\documentclass[sigplan,screen]{acmart}
\copyrightyear{2024}
\acmYear{2024}
\setcopyright{rightsretained}
\acmConference[ASPLOS '24]{29th ACM International Conference on
Architectural Support for Programming Languages and Operating Systems,
Volume 2}{April 27-May 1, 2024}{La Jolla, CA, USA}
\acmBooktitle{29th ACM International Conference on Architectural Support for
Programming Languages and Operating Systems, Volume 2 (ASPLOS '24), April
27-May 1, 2024, La Jolla, CA, USA}
\acmDOI{10.1145/3620665.3640359}
\acmISBN{979-8-4007-0385-0/24/04}

\usepackage[normalem]{ulem}
\usepackage{amsmath,amsfonts}
\usepackage[ruled,vlined,linesnumbered]{algorithm2e}
\usepackage{graphicx}
\usepackage{subfigure} 
\usepackage{tabularx}
\usepackage{float}
\usepackage{xcolor} 
\usepackage{soul} 
\usepackage{tikz} 
\DeclareRobustCommand\encircle[1]{\tikz[baseline=(char.base)]{\node[shape=circle,fill,inner sep=0.6pt] (char) {\textcolor{white}{#1}}}}
\input{utility/macro}
\setul{0.5ex}{0.3ex}
\setulcolor{purple}

\begin{document}

\title{CIM-MLC: A Multi-level Compilation Stack for Computing-In-Memory Accelerators}

\author{Songyun Qu}
\authornote{Both authors contributed equally to this research.}
\author{Shixin Zhao}
\authornotemark[1]
\affiliation{%
    \institution{Institute of Computing Technology, CAS, University of Chinese Academy of Sciences}
  \city{Beijing}
  \country{China}
  \postcode{100190}
}
\email{qusongyun18z,zhaoshixin22s@ict.ac.cn}

\author{Bing Li}
\affiliation{%
  \institution{Capital Normal University}
  \city{Beijing}
  \country{China}}
\email{bing.li@cnu.edu.cn}

\author{Yintao He}
\affiliation{%
  \institution{Institute of Computing Technology, CAS, University of Chinese Academy of Sciences}
  \city{Beijing}
  \country{China}
}
\email{heyintao19z@ict.ac.cn}

\author{Xuyi Cai}
\affiliation{%
  \institution{Institute of Computing Technology, CAS, University of Chinese Academy of Sciences}
\city{Beijing}
  \country{China}}
\email{caixuyi18s@ict.ac.cn}

\author{Lei Zhang}
\affiliation{%
  \institution{State Key Lab of Processors, Institute of Computing Technology,  Chinese Academy of Sciences}
  \city{Beijing}
  \country{China}}
\email{zlei@ict.ac.cn}

\author{Ying Wang}
\authornote{Corresponding author.}
\affiliation{%
  \institution{State Key Lab of Processors, Institute of Computing Technology,  Chinese Academy of Sciences}
  \city{Beijing}
  \country{China}}
\email{wangying2009@ict.ac.cn}

\begin{abstract}
In recent years, various computing-in-memory (CIM) processors have been presented, showing superior performance over traditional architectures. 
To unleash the potential of various CIM architectures, such as device precision, crossbar size, and crossbar number, it is necessary to develop compilation tools that are fully aware of the CIM architectural details and implementation diversity.
However, due to the lack of architectural support in current popular open-source compiling stacks such as TVM, existing CIM designs either manually deploy networks or build their own compilers, which is time-consuming and labor-intensive. 
Although some works expose the specific CIM device programming interfaces to compilers, they are often bound to a fixed CIM architecture, lacking the flexibility to support the CIM architectures with different computing granularity. 
On the other hand, existing compilation works usually consider the scheduling of limited operation types (such as crossbar-bound matrix-vector multiplication). Unlike conventional processors, CIM accelerators are featured by their diverse architecture, circuit, and device, which cannot be simply abstracted by a single level if we seek to fully explore the advantages brought by CIM.

Therefore, we propose \name, a universal multi-level compilation framework for general CIM architectures. In this work, we first establish a general hardware abstraction for CIM architectures and computing modes to represent various CIM accelerators.
Based on the proposed abstraction, \name can compile tasks onto a wide range of CIM accelerators having different devices, architectures, and programming interfaces. 
More importantly, compared with existing compilation work, \name can explore the mapping and scheduling strategies across multiple architectural tiers in CIM, which form a tractable yet effective design space, to achieve better scheduling and instruction generation results.
Experimental results show that \name achieves 3.2$\times$ inference speedup on average compared to prior CIM-oriented compilation work. 
Opensource website \url{https://cimmlc.github.io/}
\end{abstract}
\maketitle 

\section{Introduction}
\label{intro}
Advancements in computer performance have been trapped by the famous ``\textit{memory wall}'' problem for decades~\cite{wulf1995hitting}. 
The advent of data-intensive workloads, like DNNs~\cite{liu2017survey}, further exacerbates this problem.
As a promising technology to combat the memory wall, computing-in-memory (CIM) has garnered significant attention for its capability of reducing frequent data movement and parallelizing multiply-and-accumulate (MAC) operations. 
A variety of CIM-based architectures emerge, achieving significantly improved computing efficiency for DNN applications over traditional architectures~\cite{ankit2019puma,biswas2018conv, chen2020survey,chi2016prime,eckert2018neural,Shafiee2016isaac,song2017pipelayer}.

To fully exploit CIM, an efficient compilation tool that bridges the DNN scope and various CIM hardware is needed. However, the development of CIM accelerators shows the following trends that pose thorny challenges to the design of a general compilation tool:
\textbf{(1) The diversity of CIM memory devices.} 
Unlike the traditional computing paradigm, CIM hinges on memory devices to perform computations, in which the attributes of memory devices, like their types and precision levels, exert a substantial influence on the feasible scheduling space within a CIM compilation framework.
For example,  when comparing SRAM\cite{he2022processing} and ReRAM, although both have similar latency for read operations, the cost of writing data is considerably higher in ReRAM~\cite{angizi2019accelerating}. 
Thus, SRAM-based CIM supports flexible data read and write updates on CIM memory~\cite{biswas2018conv}, while ReRAM-based CIM usually assumes that weights are frozen in the crossbar, avoiding the penalty of frequent writes~\cite{chi2016prime,Shafiee2016isaac}. 
These factors seriously impact the scheduling strategy and space of a CIM compiler.

\textbf{(2) The diversity of CIM architectures.} 
In contrast to the common tiled PE-arrays of typical DNN accelerators~\cite{chen2016eyeriss}, the architecture of CIM exhibits a diverse and distinct organization\cite{li2019overview}, like the structure of an individual computing unit, the number of crossbars within each computing unit, crossbar size, \etc.
Consequently, the optimization space of the CIM compiler becomes much larger. The CIM compiler must possess an understanding of the manifold forms within CIM architecture when scheduling and mapping DNN operations onto CIMs.

\textbf{(3) The diversity of CIM programming interface.} 
CIM architectures have rich hierarchies, and the manner in which they are exposed to programmers varies across different chip designs.
For example, some CIMs support fined-grained row-wise operations to realize more general-purpose arithmetic except convolutions and matrix multiplication~\cite{jain2021wlm}.
To fully exploit CIM architecture across different applications, researchers have developed the CIM-oriented programming interface~\cite{ankit2019puma,fujiki2018memory}. 
However, the existing programming interface is deeply bound to a specific CIM design. 
With different programming interfaces, the granularity of the elementary computing unit that the scheduler can manage during optimization varies. A fine granularity leads to a more complex scheduling space, which increases the difficulty of compilation optimization.

Therefore, the key challenges to developing a general-purpose CIM compiler are: \textit{1.~How to perform multi-dimensional hardware abstraction on CIM to enhance the generality of the compiler?}  \textit{2.~How to generate the optimal mapping and scheduling for different CIMs that expose different operation levels?}

Unfortunately, existing works are difficult to adapt to the diversity of CIM accelerators, and lack optimization considerations for the specific computing granularity in different CIMs, resulting in inferior performance and even failure of compiling. 
Some works manually deployed the model on CIMs~\cite{Shafiee2016isaac} with customized mapping and scheduling policies that are hard to generalize to other CIMs.
Some works proposed CIM-oriented compilation tools~\cite{ambrosi2018hardware, drebes2020tc, han2021polyhedral}. However, they are insufficient to tackle the above challenges. 
Most of the CIM compilers neglect the generality in their design. 
For example, Ambrosi \etal~\cite{ambrosi2018hardware} propose a compilation tool that schedules matrix-vector computation (MVM) on a ReRAM-based architecture, but its performance degrades when the CIM architecture and computing granularity change.
Meanwhile, they may also fail to fully explore the scheduling space of DNN operators at the corresponding computing level. 
For example, Han \etal~\cite{han2021polyhedral} design a compilation tool to deploy DNNs on the ISAAC architecture~\cite{Shafiee2016isaac} via an MVM-grained programming interface, but its optimization strategy stays at the computing graph level 
neglecting the opportunity to fine-control the crossbar resource allocation and MVM operation sequencing for better results.

In this work, we propose a multi-level compilation stack for CIM accelerators to achieve more versatile and efficient compilation.  
In particular, we first introduce a three-tier architecture abstraction of CIM hardware, from the crossbar tier to the chip tier. 
Each tier has corresponding hardware architecture parameters, which can be designated to describe a particular CIM accelerator.
Then, based on the hardware abstraction of CIM accelerators, we propose the computing mode abstraction with computing granularity ranging from fine to coarse to suit various programming interfaces.
At different computing modes, we abstract the basic operation types of a CIM accelerator as meta-operator sets, specifically supported operators in the CIM like memory row read/write, and MVM operator with crossbars. 
We can then use the meta-operator sets to represent the computation process of DNNs on CIM accelerators.

Within the provided software and hardware space, we propose a multi-level scheduling approach to optimize mapping and scheduling for different computing modes. According to the computing mode given by the target CIM, our policy progressively optimizes computations from coarse-level computing graphs to fine-level vector computing operations and generates the corresponding meta-operator flow. 
Compared to existing compilation methods that optimize schedules at a single computational level, such as the DNN computing graph, our approach offers a more holistic optimization perspective, resulting in higher computing efficiency.

The contributions of this work are summarized as follows:
\begin{itemize}
\item This paper proposes a compilation stack, \names, which enables automatic compilation for various CIM accelerators. In \names, we propose a hardware abstraction of hierarchical architecture and computing mode of CIM architecture to support a wide range of CIM accelerators.
\item Then, we propose a multi-level DNN scheduling approach, which realizes the flexible DNN operators mapping and scheduling based on the computing granularity exposed by specific CIM accelerators. We conduct thorough optimization at each abstraction level to generate the operation flow. The proposed multi-level optimization flow covers a much broader scheduling space than the previous graph or MVM-level scheduling but also avoids the intractability faced by single-level fine-grained scheduling.
\item Compared with existing compilation work~\cite{han2021polyhedral}, the proposed \name can improve the inference speed by up to $3.2\times$. We also verify the proposed compilation stack on existing CIM accelerators, PUMA~\cite{ankit2019puma}, CIM work~\cite{jain2021wlm} and CIM work~\cite{jia202115}, to demonstrate the generality of \names. Experiments show that \name can accelerate the inference
speed of CIM work~\cite{jain2021wlm} and CIM work~\cite{jia202115} by $2.3\times$ and $3.7\times$, respectively, while reducing the peak power consumption by 75\% for PUMA~\cite{ankit2019puma}.

\end{itemize}

\section{Background and Motivation}

\subsection{The Diversity of CIM for DNNs}
\begin{figure}[h]
    \centering
    \includegraphics[width=\linewidth]{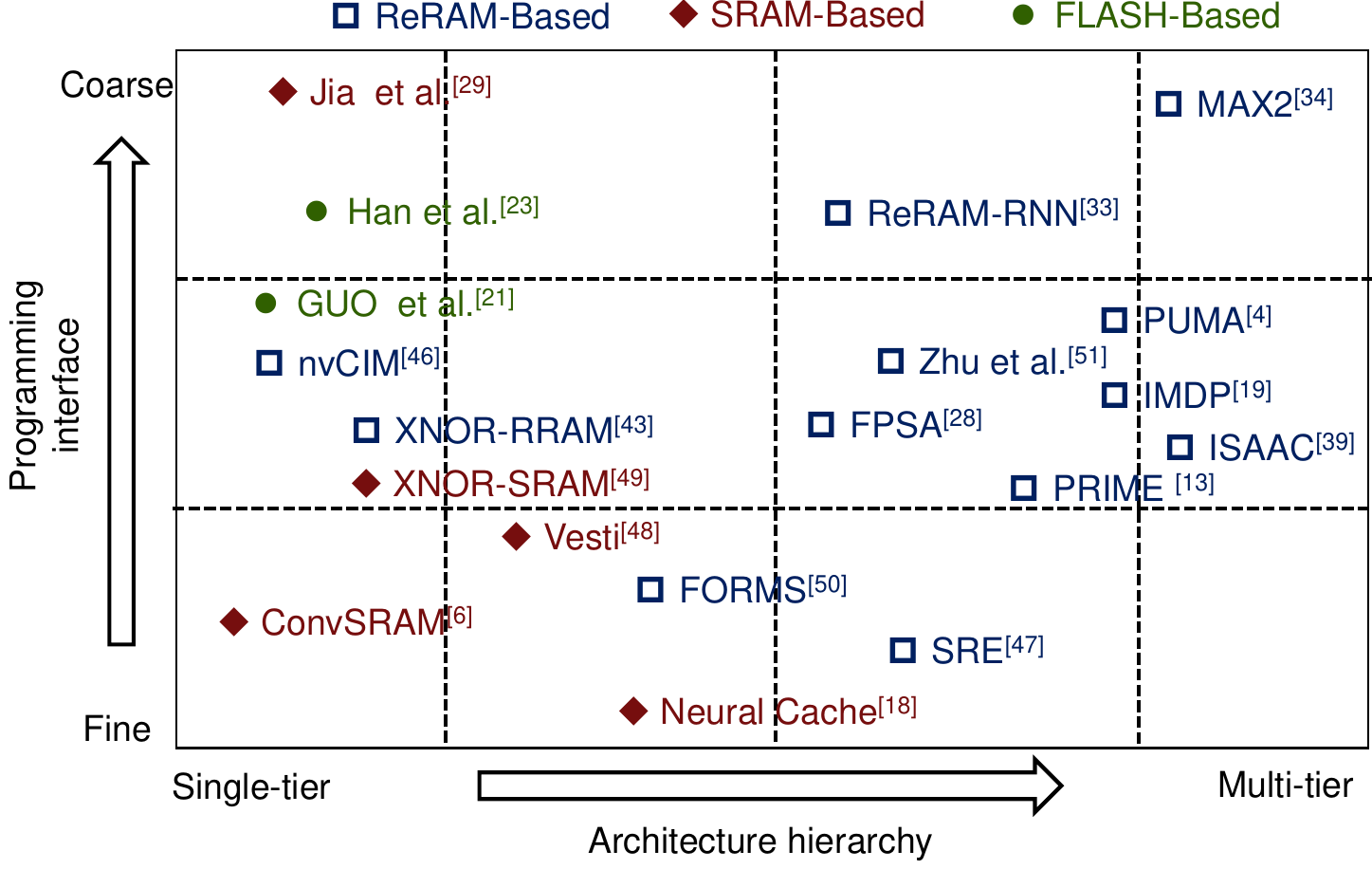}
    \caption{Diversity of CIM architecture.}
    \label{fig:background_1}
\end{figure}

\begin{table*}[ht]
\small
\centering
\caption{Comparison of the generality of this work and the existing works.}
\resizebox{1.0\textwidth}{!}{
\begin{tabular}{@{}cccccccc@{}}
\toprule
 & \multicolumn{3}{c}{Supported Device Type } & \multicolumn{3}{c}{Supported Programming Interface} & Optimization Granularity \\ \midrule
 & SRAM & ReRAM & 
MISC(\footnotesize{\eg PCM, FLASH)}
 & VVM & MVM & DNN Operators & / \\
PUMA \cite{ambrosi2018hardware,ankit2019puma} & × & \checkmark & / & × & \checkmark & × & MVM\\
IMDP \cite{fujiki2018memory} & × & \checkmark & / & \checkmark & \checkmark & × & MVM \\
TC-CIM \cite{drebes2020tc} & × & \checkmark & / & × & \checkmark & × & MVM \\
Polyhedral-based \cite{han2021polyhedral} & × & \checkmark & / & × & \checkmark & \checkmark &MVM, MM,  Conv \\
OCC \cite{siemieniuk2021occ} & \checkmark & \checkmark & / & \checkmark & \checkmark & × & / \\
Ours & \checkmark & \checkmark & \checkmark & \checkmark & \checkmark & \checkmark & VVM, MVM, DNN Operators \\ 
\bottomrule
\end{tabular}}
\label{sec4T1}
\end{table*}

Researchers have proposed various CIM-based DNN accelerators. We sort out the designs of recent CIM accelerators from three dimensions: memory device, architecture hierarchy, and programming interface, and summarize them in \fig\ref{fig:background_1}~\cite{ankit2019puma,biswas2018conv,chi2016prime,eckert2018neural,fujiki2018memory,guo2017fast,han2019novel,ji2019fpsa,jia202115,long2018reram,mao2019max,Shafiee2016isaac,sun2018xnor,xue202116,yang2019sparse,yin2019vesti,yin2020xnor,yuan2021forms,zhu2019configurable}. 
SRAM, ReRAM, and FLASH are included in the dimension of the memory device. Their different read/write latency and storage density affect the data mapping and scheduling policy of these CIMs. For example, the write latency of ReRAM/FLASH is relatively long, so CIMs based on ReRAM/FLASH often ford write operations during computation (\eg Guo \etal~\cite{guo2017fast} and PRIME~\cite{chi2016prime}).
As for the architecture hierarchy, some CIM designs hierarchically organize tiles, cores, and crossbars from top to bottom (\eg ISSAC~\cite{Shafiee2016isaac} and MAX2~\cite{mao2019max} ) while others do not (\eg a single-tier hierarchy only having tiled crossbars in ConvSRAM~\cite{biswas2018conv} and nvCIM~\cite{xue202116}). Additionally, even though ISAAC and PUMA \cite{ankit2019puma,Shafiee2016isaac} have a similar hierarchical structure, they still use different crossbar sizes and the number of crossbars. 
Accordingly, their model mapping ways are different. The scheduling space during the compilation is different as well.
The programming interface refers to the computing granularity that CIM can support from the users' view. As for the coarse-grained programming interface, the DNN computation is decomposed into convolution operations and then is performed by CIM cores (\eg Jia \etal's work~\cite{jia202115}), while for the fine-grained programming interface, the bit-wise vector operations are supported by CIM crossbars (\eg ConvSRAM~\cite{biswas2018conv}). Other works propose to decompose DNN operators and schedule MVM-grained operation in CIM crossbars~\cite{ankit2019puma,fujiki2018memory,sun2018xnor,yin2020xnor,qu2022coordinated}. 
To sum up, the diversity of existing CIMs for DNN raises a growing demand for an efficient general compilation tool.

\subsection{Compilation Tools for CIM}
\begin{figure}[!tbp] 
    \centering
    \includegraphics[width=\linewidth]{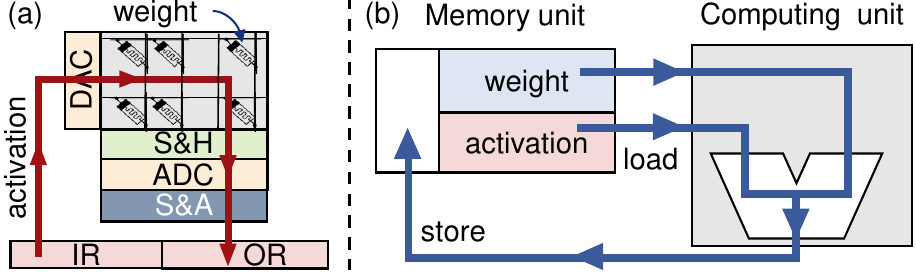}
    \caption{Computation dataflow comparison of (a) CIM and (b) traditional architecture.}
    \label{fig:dataflow}
\end{figure}

Neural network compilers for different computing architectures have been widely discussed in prior works, such as TVM~\cite{chen2018tvm}.
However, existing machine learning compilers primarily focus on compilation optimization related to traditional architecture hardware where the memory and computing units are separated~\cite{cyphers2018nGraph,rotem2018glow,sabne2020xla,vasilache2018tensor,cai2022optimus,chen2022framework,gao2023layer}.
As \fig \ref{fig:dataflow} shows, traditional architecture and CIM are
totally different computation paradigms.
As for traditional architectures being bottlenecked by memory accesses, their compilation tools are composed of optimization methods for memory efficiency improvement.
For instance, the popular loop unrolling and unfolding approaches in the traditional compilation tools are for improving data locality~\cite{li2020deep}. 
In contrast, CIM architectures performing computations inside memory~\cite{Shafiee2016isaac,he2021tare} shall focus on improving the efficiency of in-situ computation instead of improving the memory efficiency in the compilation techniques and hence desire a specific compilation tool. 

\begin{figure*}
    \centering
    \includegraphics[width=\linewidth]{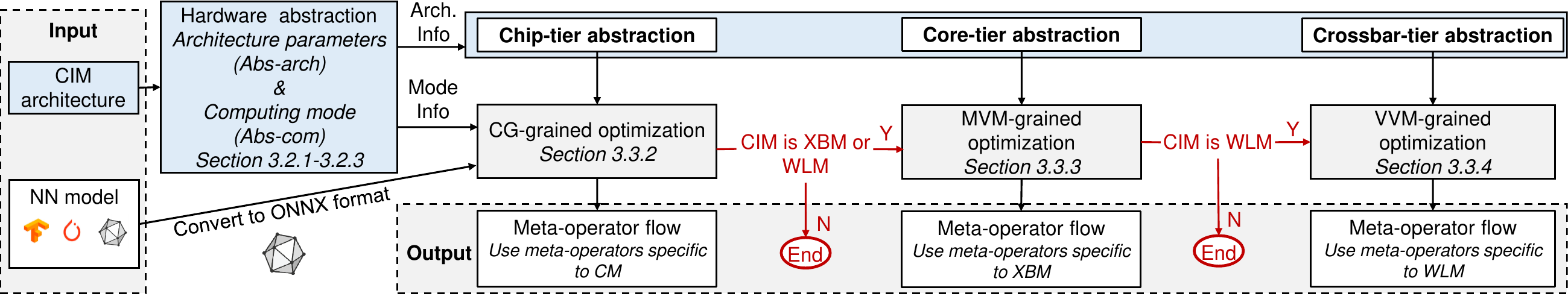}
    \caption{Overall workflow of CIM-MLC.}
    \label{new_overview}
\end{figure*}
Some researchers have noticed the problem and developed some CIM compilation tools~\cite{drebes2020tc,han2021polyhedral}. However, we observe these works are inadequate as a general compilation tool for CIM when summarizing them from the dimensions of the device type, programming interface, and optimization granularity (as shown in \tab\ref{sec4T1}).

Considering the device types involved, PUMA ~\cite{ambrosi2018hardware,ankit2019puma} and IMDP~\cite{fujiki2018memory} present compilation methodologies tailored to particular CIM architectures that rely on ReRAM. However, these approaches lack interoperability with alternative devices and programming interfaces.
A few works have explored general compilation frameworks. 
TC-CIM~\cite{drebes2020tc} and Polyhedral-based~\cite{han2021polyhedral} harness polyhedral models to facilitate the compilation process on CIM accelerators, which automatically identify and map MVM operations inherent in DNNs onto CIM.   
However, these two works also mainly focus on ReRAM-based CIM and assume that there are ample memory resources available for parameter loading or duplication, which overlooks crucial facts of resource-constrained situations that are typically encountered in SRAM-based CIMs.
Relatively, OCC~\cite{siemieniuk2021occ} is a comprehensive compilation that encompasses abundant device types as well as numerous programming interfaces. Built upon a specialized MILR with an ISA, this work enables the optimization at various levels of granularity.
Nonetheless, this work does not incorporate the coarser-grained programming interface like the DNN operator and does not explore the mapping optimization of DNN computation on CIMs. 
To summarize, current research on CIM compilers lacks sufficient abstraction of hardware and is limited in support for various architectures and programming interfaces in CIM accelerators. 

Regarding optimization granularity, most of the works~\cite{ambrosi2018hardware,ankit2019puma,drebes2020tc,han2021polyhedral} support the mapping and optimization of DNN at the level of MVM operations. However, due to the increased complexity of the scheduling space, existing work has not fully explored the optimization possibilities at the MVM granularity. For example, some of them~\cite{ambrosi2018hardware,ankit2019puma,han2021polyhedral}
support the pipeline inter-network layers but do not consider the computing pipeline opportunity at the MVM-grained. 
In fact, optimizing the scheduling of MVM operations helps to further optimize computing efficiency.

\vspace{-12pt}
\section{Methodology}
\subsection{Overview}
Figure~\ref{new_overview} demonstrates the overview of the proposed CIM-MLC.
Existing CIM compilation works have poor generality because of their single-level scheduling policy, 
while CIM-MLC is a general compiler that features unified abstraction from diverse hardware and multi-level scheduling with abundant meta-operators. 
The hardware abstraction provides the same description interface of architecture parameters (Abs-arch) and computing mode (Abs-com) to various CIM designs.
To decouple the data mapping and computing scheduling with one architectural design, we propose the multi-level scheduling technology to handle the computing mode for different architectural tiers in CIMs. 
The multi-level scheduler tailors the optimization method for each computing mode, applies the optimization method independently or jointly according to the abstraction of the CIM accelerator, and finally generates the meta-operator flow for the CIM accelerator.

\begin{figure*}[tb]
    \centering
    \includegraphics[width=.8\linewidth]{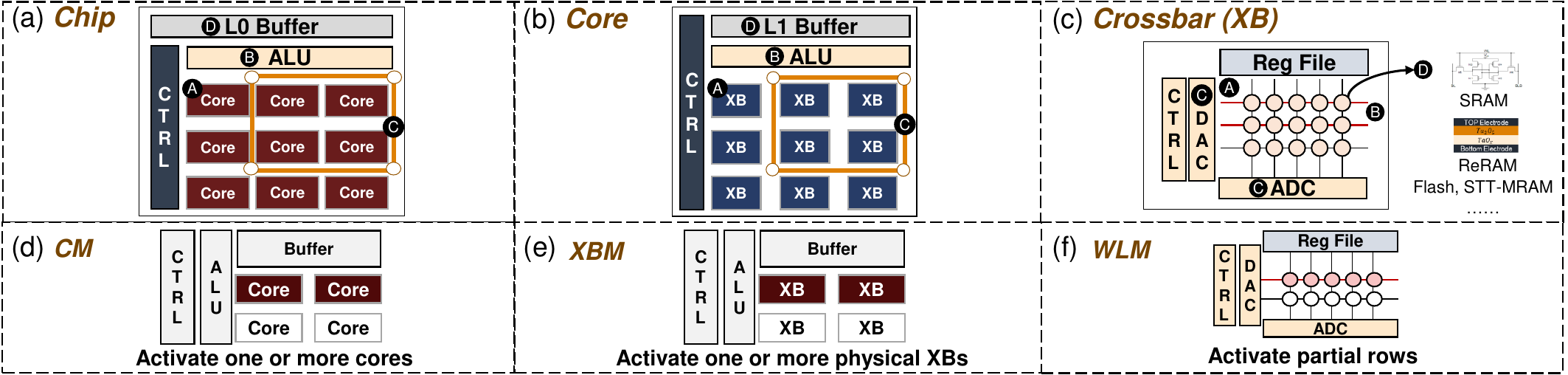}
    \caption{CIM Hardware Abstraction.}
    \label{f1}
\end{figure*}

\subsection{CIM Hardware Abstraction}
In order to support a wide range of CIM accelerators, we first construct the hardware abstraction, covering two key aspects: the architecture parameters (Abs-arch) and the computing modes (Abs-com), respectively.
We model the CIM-based DNN accelerator as a hierarchical architecture, which contains three tiers from top to bottom: (a) chip, (b) core, and (c) crossbar. Meanwhile, similar architectures may support different levels of CIM operation granularity. Some CIM designs only support specific operators or fixed-granularity MVM computation, while others offer interfaces for controlling rows, enabling a wide range of computation. So, we design a three-level computing abstraction for CIM. As illustrated in \fig \ref{f1}(d)-(f), the top-to-bottom levels are Core Mode (CM), Crossbar Mode (XBM), and Wordline Mode (WLM), denoting the coarse-grained to fine-grained DNN operators, respectively.
Architecture abstraction tiers and computing mode abstraction levels maintain a one-to-one correspondence. The hardware scheduling granularity provided by the CIM architecture determines the supported computing mode and the architecture abstraction parameters exposed to the compiler. We also develop a comprehensive set of meta-operators, which facilitate compilation scheduling optimization and instruction generation and enable users to define and customize hardware-supported operations within our framework.

\begin{figure}[tp]
    \centering
    \includegraphics[width =\linewidth]{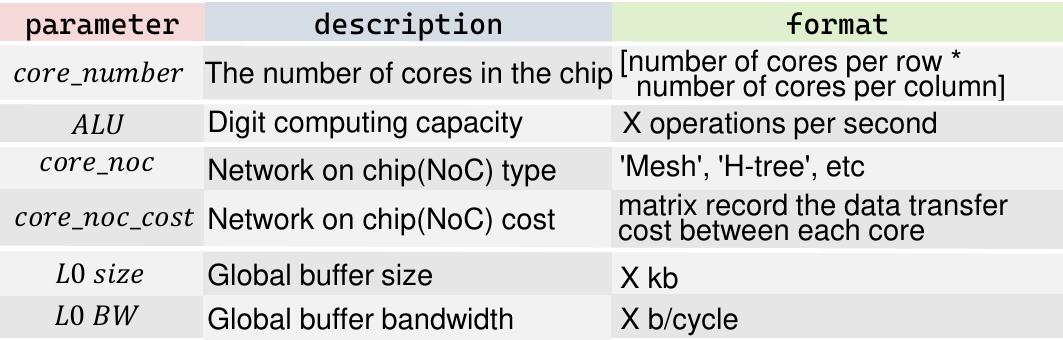}
    \caption{Chip tier architecture abstraction parameters.}
    \label{fig:p3_chiptier}
\end{figure}

\begin{figure}[tp]
    \centering
    \includegraphics[width = \linewidth]{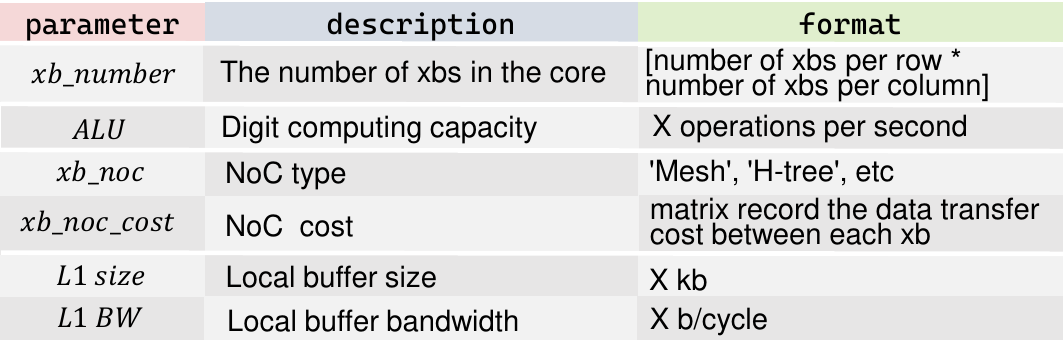}
    \caption{Core tier architecture abstraction parameters.}
    \label{fig:p3_coretier}
\end{figure}

\subsubsection{Chip tier architecture abstraction and core mode} 
As shown in the \fig\ref{f1}(a), in the chip tier, multiple cores (\encircle{A}) connected via a network-on-chip (NoC \encircle{C}) are grouped into a single chip, each used for operators like convolution. A shared on-chip memory (\encircle{D}) is used for data storage, and digital arithmetic logic units (ALU \encircle{B}) are used for computations beyond the scope of CIM-supported operators.

At this tier, the minimum scheduling granularity provided by the CIM architecture is core. We abstract its computing mode as core mode (CM). In this mode, the compiler assigns one or more cores to complete one DNN operator (\eg convolution) according to the operator's demand and the core's capability. The scale of computation supported by a single core and the total number of cores determine the maximum number of operators that can be concurrently mapped on a chip. Hence, we employ the parameter \texttt{\textbf{core\_number}} to record the total number of cores within the chip, thus abstracting the chip's CIM computational capacity.

The ALU performs commonly used neural network operators such as activation functions, pooling, and CIM-specific operations like shift-accumulate. ALU's functionality and computational speed impact the compilation of digital computation operators. We use \texttt{\textbf{ALU}} to denote the ALU computation speed and the functionality will be recorded by meta-operator, which we will show in Section 3.3.2.

The type of on-chip network and data transfer rate determine the data transfer efficiency between cores. We use \texttt{\textbf{core\_noc}} and \texttt{\textbf{core\_noc\_cost}} to abstract the on-chip network. Meanwhile, as the storage capacity and bandwidth of the global buffer jointly determine the execution time of data movement, \texttt{\textbf{L0 size}} and \texttt{\textbf{L0 BW}} are employed to record on-chip buffer capacity and bandwidth.

The architecture parameters in chip tier are summarized in the \fig \ref{fig:p3_chiptier}. We use the above mentioned parameters to abstract the CIM architecture characteristics in the chip tier and expose the parameters to the compilers in CM. CIM-MLC will combine the characteristics of operators in DNN and CIM architecture parameters to optimize latency and energy consumption during operator mapping.

\subsubsection{Core tier architecture abstraction and crossbar mode}
As shown in the \fig\ref{f1} (b), the core tier abstraction includes the features within a core, which consists of multiple crossbars (\encircle{A}), local buffers (\encircle{D}), and digital computation units (\encircle{B}). The crossbars are connected via NoC (\encircle{C}).

The minimum scheduling granularity at this tier is crossbar, which we abstract as crossbar mode(XBM). One or multiple crossbars work together to complete one matrix-vector multiplication in this mode. Similar to chip tier, we use the \texttt{\textbf{xb\_number}} to record the number of crossbars that work at the same time, which can decide the maximum scale of matrix-vector multiplications that can be concurrently mapped on the core. The convolution operator in DNN is decomposed into a sequence of matrix-vector multiplications in the XBM.

\begin{figure}[tb]
    \centering
    \includegraphics[width=\linewidth]{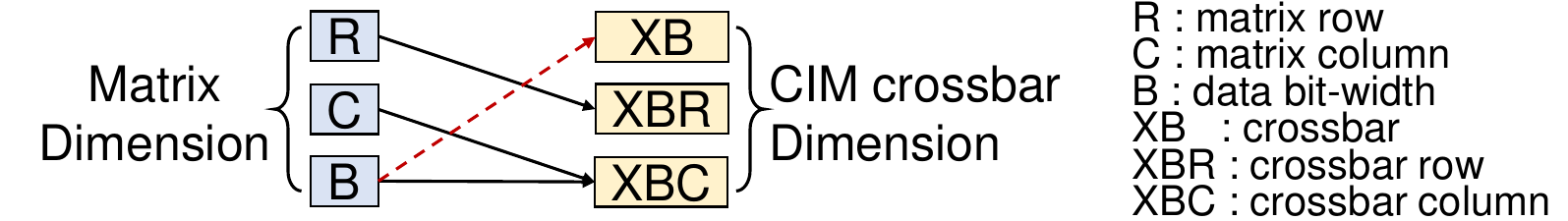}
    \caption{Dimension binding for DNNs mapping on CIM crossbar.}
    \label{figxb1}
\end{figure}

\begin{figure}[tp]
    \centering
    \includegraphics[width = \linewidth]{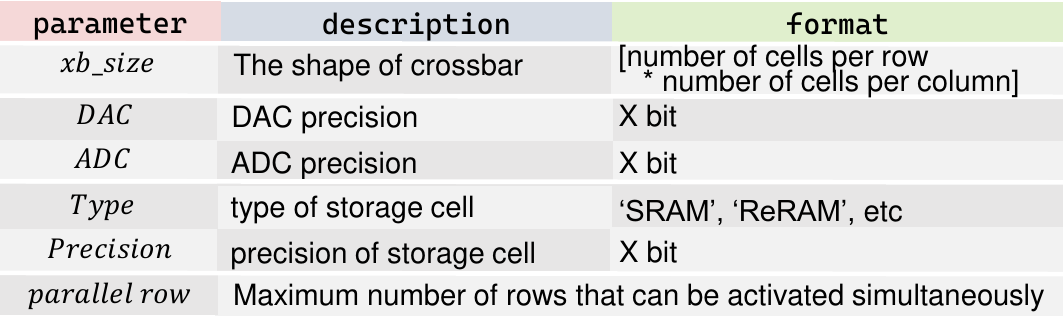}
    \caption{Crossbar tier architecture abstraction parameters.}
    \label{fig:p3_crossbartier}
\end{figure}

To accommodate different CIM designs for executing MVM on crossbars~\cite{ankit2019puma,Shafiee2016isaac,song2017pipelayer,zhu2019configurable}, we introduce the concept of \textit{VXB (Virtual Crossbar)} as the computational unit rather than physical crossbars to facilitate the computing scheduling in the compiler. A dimension-binding scheme is designed to construct a VXB, specifying which crossbars collaborate to perform a single MVM.
\fig\ref{figxb1} illustrates the dimension-binding scheme, where the matrix dimension has matrix row (R), column (C), and data bit-width (B). The computing crossbar dimension refers to the crossbar itself (XB), crossbar row (XBR), and column (XBC). \fig\ref{figxb1} shows that the matrix dimension R/C/B is bound to the XBR/XBC/XBC, respectively. This binding represents that the data bits are spread to the adjacent column in the crossbar. If the matrix dimension B is bound to the XB, the data bits will be spread to the different crossbars.

Similar to chip tier, we utilize \texttt{\textbf{ALU}} to denote the speed of the digital computation unit, \texttt{\textbf{xb\_noc}} and \texttt{\textbf{xb\_noc\_cost}} for NoC abstraction, and \texttt{\textbf{L1 size}} and \texttt{\textbf{L1 BW}} to represent the local buffer characteristics at the core tier.

The architecture parameters in core tier are summarized in the \fig \ref{fig:p3_coretier}. These chip and core tier parameters are exposed to the compilers in XBM as core tier abstraction records the architecture details within the core, and chip tier parameters record the whole top tier architecture. CIM-MLC leverages these parameters to map operators to crossbars, optimizing the crossbar utilization.

\subsubsection{Crossbar tier architecture abstraction and wordline mode}

As shown in the \fig\ref{f1} (c), The crossbar tier abstraction serves as the fundamental computational unit, describing the details component within the crossbar. The crossbar tier has a memory crossbar (\encircle{A}) with its peripheral circuits (\eg wordline drivers, bitline drivers, and signal converters (ADC \encircle{C}, DAC \encircle{C}, etc.)). Each crossbar array's rows (\encircle{B}) can work independently.

At this tier, the minimum scheduling granularity provided by the CIM architecture is rows. We abstract this as wordline mode (WLM) and use \texttt{\textbf{parallel row}} to record the number of rows that can be activated concurrently. The mode allows the CIM to optimize the power cost or alleviate the variation by closing partial rows of one crossbar~\cite{jain2021wlm,yang2019sparse}. Users can define the number of rows that can be activated simultaneously in a crossbar. Meanwhile, the shape of the crossbar determines the maximum scale of matrix-vector multiplication calculations. So, we use the parameter \texttt{\textbf{xb\_size}} to record the size of a crossbar. 

The attributes of memory cell (\encircle{D}) within the crossbar profoundly impact the computation process. Additionally, memory cell precision influences data representation and the requisite number of crossbars for operators, affecting scheduling decisions. Hence, we use the parameters \texttt{\textbf{Type}} and \texttt{\textbf{Precision}} to respectively record the memory cell type and precision.

For the peripheral circuits, the Analog-to-Digital Converter (ADC) or a special sense amplifier usually performs pre-defined calculations on the readout analog signals from the crossbar. The Digital-to-Analog Converter (DAC) is also needed to complete data signal conversion. The precision of DAC and ADC influences computation accuracy and latency. We use parameters \texttt{\textbf{DAC}} and \texttt{\textbf{ADC}} to record the precision of DAC and ADC.

The architecture parameters in crossbar tier are summarized in the \fig \ref{fig:p3_crossbartier}. We expose the whole three tiers of architecture abstraction parameters to the compiler in WLM. CIM-MLC will transform the computing process in DNNs into the row-wise read and write of the CIM crossbar.

\subsection{Multi-Level Scheduling}
With the architecture abstraction and computing mode, we apply a multi-level scheduling strategy to optimize the efficiency of DNN inference on CIM, achieving low-latency, high energy-efficient deployment of models.
For each computing mode abstraction in the associated abstracted architecture tier, we design the optimization method. 

Next, we introduce the whole workflow and the detailed optimization methods in the multi-level scheduling strategy.

\subsubsection{Multi-Level Scheduling workflow}
The compilation process is illustrated in \fig \ref{new_overview}. Firstly, the compiler gets the DNN models in ONNX format \cite{bai2019}. ONNX represents the model by computation graph, in which nodes correspond to operators, and edges denote the data dependency between each operator. The compiler progresses from coarse to fine granularity, optimizing the computations in Computational Graph Grained (CG-Grained), Matrix-Vector Multiplication Grained (MVM-Grained), and Vector-Vector Multiplication Grained (VVM-Grained). These optimizations are tailored to the CM, XBM, and WLM architectures. We adopt a multi-level joint scheduling optimization strategy to explore the scheduling space of the CIM architecture effectively.

When targeting the CM architecture, the compiler performs scheduling optimization at the CG-grained only. CG-grained optimization aims to explore operator duplication and pipeline strategies while considering resource constraints. The compiler takes the ONNX model and chip-tier hardware parameters as inputs. The optimization information for each operator is recorded by adding attributes to the nodes in the ONNX graph, such as the operator's duplication.
For the XBM architecture, the compiler inherits the optimization results from the CG-grained, and chip and core-tier hardware parameters to perform the MVM-grained optimization. MVM-grained optimization further explores the finer operator duplication and pipeline before mapping operators to the crossbars.
When dealing with the WLM mode architecture, the compiler builds upon the optimizations from the CG-grained and MVM-grained and, at a finer granularity of row scheduling, performs optimization at the VVM-grained. 

\begin{figure}
    \centering
    \includegraphics[width=\linewidth]{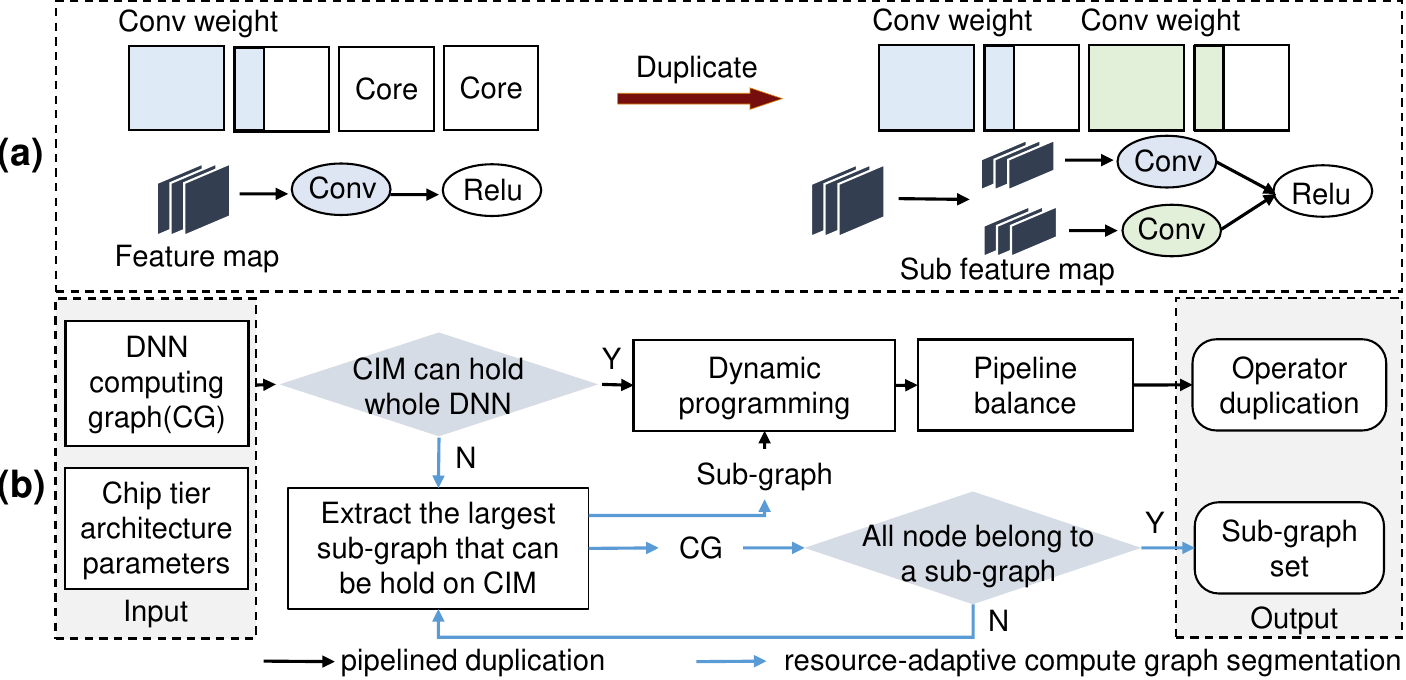}
    \caption{CG-grained optimization (a) Operator duplication illustration (b) Optimization algorithm.}
    \label{fig:p3_cg}
\end{figure}

\subsubsection{CG-Grained Optimization}

In the core mode, the compiler maps DNN operators onto cores following the chip tier architecture parameters. We propose a computing graph-grained optimization approach to improve operator mapping, 
all without altering the operator execution process. The primary objective of this optimization level is to judiciously utilize hardware resources, thereby reducing latency and power consumption.

Current graph-level optimization technology often overlook hardware characteristics \cite{chen2018tvm}. We introduce a novel CG-grained optimization, exploring the \textit{duplication} scheme and \textit{pipeline} of DNN operators on CIM cores under the total \textbf{\texttt{core\_number}} constraint. As shown in the \fig \ref{fig:p3_cg} (a), the operator duplication strategy enhances computational throughput by appropriately duplicating CIM-supported operators on cores. Meanwhile, the inter-operator pipeline strategy seeks to enhance execution efficiency.

Considering various hardware resource constraints of the CIM architecture, we propose a resource-adaptive compute graph segmentation and intra-segment dynamic balancing pipelined duplication algorithm as \fig\ref{fig:p3_cg} (b) illustrated. This algorithm takes the ONNX format computation graph and chip-tier hardware parameters as input and outputs the subgraph set of the computing graph and the duplication results for CIM-supported operators.

First, we initialize the resources and computation latency of each node. For convolution nodes, the computation latency scales with the output feature map size. We use dynamic programming to search for all operators' duplication numbers under the \textbf{\texttt{core\_number}} constraint. Then, to avoid the pipeline stall because of the imbalance between computing and data access of adjacent layers, we adjust the duplication number for each node. Meanwhile, considering the constraint of \textbf{\texttt{core\_noc\_cost}} and \textbf{\texttt{L0 BW}}, CIM-MLC will update the duplication number to keep the data transfer amount within the NOC and buffer capability. Once the CIM-unsupported node, like Relu, follows the operator, we will also update the duplication number under the constraint of \textbf{\texttt{ALU}}.
\begin{figure}[tb]
    \centering
    \includegraphics[width=\linewidth]{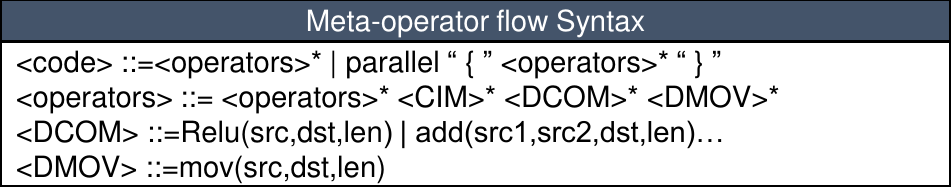}
    \caption{The syntax of code generation in Backus Naur Form (BNF).}
    \label{fig:p3_syntax_all}
\end{figure}

\begin{figure}[tb]
    \centering
    \includegraphics[width=\linewidth]{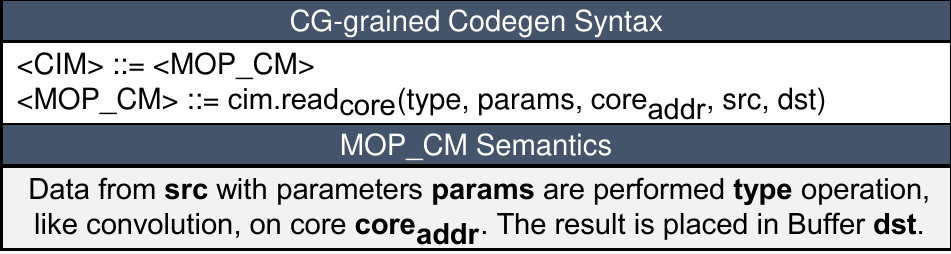}
    \caption{The syntax of CG-grained codegen in BNF format and MOP\_CM semantics.}
    \label{fig:p3_syntax_cm}
\end{figure}
As for the case that CIM resources are not able to hold a whole DNN, we iteratively construct the maximal sub-graphs that can fit within CIM capacity. 
For each constructed computation sub-graph, we update it by successively popping the last nodes from the subgraph. Then, we use the dynamic programming algorithm to get the latency of the remaining computation sub-graph after a node is popped out. Once the latency no longer decreases, the construction of that sub-graph is complete.
The nodes that pop out will be used to construct the next maximal sub-graphs until all nodes are part of a single sub-graph and have duplication numbers assigned.

Ultimately, we obtain the results of computing graph segmentation and operator duplication numbers at this grained optimization, which are subsequently passed on to the next optimization level.

\textbf{Meta-operator Flow Generation} 
After the optimization, the compiler backend binds operator execution to the corresponding cores. 
As primitives used in the current ML compiler \cite{li2020deep} can not support the description of CIM, we introduce a CIM meta-operator set tailored for CM (MOP\_CM) to describe the hardware activation at the chip tier. 
As the \fig \ref{fig:p3_syntax_cm} shows, MOP\_CM includes the $CIM.read_{core}$ instruction. 
In addition to the CIM meta-operators, we have also designed corresponding meta-operators for other operations (\ie~DCOM for digit computing operations and DMOV for data movement, respectively). Users have the flexibility to extend meta operators, aligning them with the hardware-supported functions. 
The compiler will compile relevant operators based on the provided meta-operators. The simplified syntax of code generation is shown in the \fig \ref{fig:p3_syntax_all}, in which the label \texttt{parallel} indicates the operators executing in parallel.

\subsubsection{MVM-Grained Optimization}

\begin{figure}
    \centering
    \includegraphics[width=\linewidth]{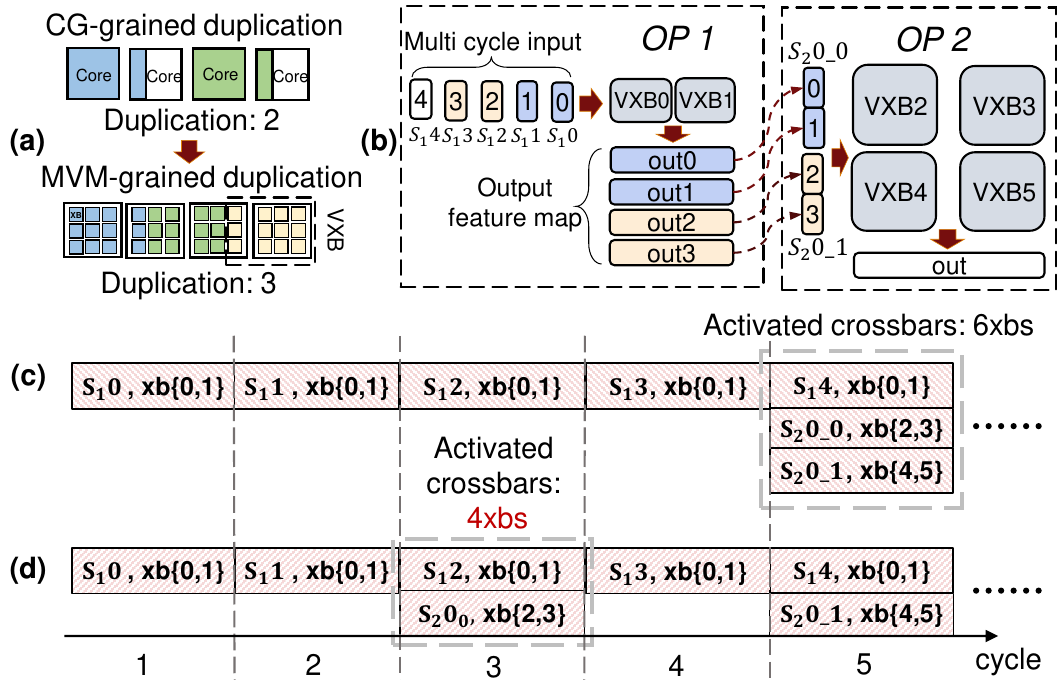}
    \caption{MVM-Grained Optimization (a) Operator duplication at VXB granularity (b) MVM-grained operator mapping and pipeline (c) Traditional pipeline activated crossbars (d) MVM-grained pipeline activated crossbars.}
    \label{fig:p3_xb}
\end{figure}

As for the XBM, the compiler unrolls the CIM-supported operator to matrix-vector multiply and then maps and schedules the MVM to crossbars.
We design MVM-grained optimization after completing CG-grained scheduling, aiming to improve resource utilization and computing throughput during the slide of weight kernels on feature maps in the convolution by effectively using the crossbars in each core.

Specifically, MVM-grained optimization explores two key techniques: the \textit{duplication} of the operator in the crossbars and MVM-grained computing \textit{pipeline} to boost the computing throughput under the power limitation. 
The compiler receives the segmentation of the computing graph enriched with CG-level optimization information and chip and core-tier hardware abstraction, completing the MVM-grained optimization under the constraints of available \textbf{\texttt{xb\_number}}.

We updated the \textit{duplication} number of an operator within crossbars (as shown in \fig \ref{fig:p3_xb} (a)), which get the duplicated number using the following Equation:
\begin{equation}\label{finedup}
\small
\begin{aligned}
& D'^{O_i} =\lfloor \frac{num^{O_i}_{core}*D^{O_i}*Core_{VXB}}{num^{O_i}_{VXB}} \rfloor,\\
\end{aligned}
\end{equation}
where $Core_{VXB}$ is the number of VXBs in each core, $D^{O_i}$ is the duplication number of the operator determined by the CG-grained optimization, $num^{O_i}_{core}$ is the number of cores occupied by this operator $O_i$, $num^{O_i}_{VXB}$ is the number of VXBs occupied by this operator $O_i$, which can be calculated based on the dimension and the size of VXB and the operator. Usually, one operator demands multiple VXBs to store its weights and complete the calculation.

MVM-grained computing \textit{pipeline} strategically staggers the activation time of different crossbars to reduce peak power consumption as illustrated in \fig\ref{fig:p3_xb} (b).
As we need to map a matrix-vector multiplication to multiple crossbars, we usually wait until all crossbars receive their inputs before computing in the traditional scheduling \cite{Shafiee2016isaac}. The pipeline strategy we propose, however, activates a crossbar as soon as it receives its input, completing the computing mapping to the crossbar. This reduces the number of crossbars that need to be activated simultaneously, thus lowering peak power consumption.

In the example, Operator 1 (OP 1) is mapped to VXB~0 and VXB~1, and Operator 2 (OP 2) is mapped to VXB 2-5.
Input $S_1i$, corresponding to one sliding window in convolution, enters OP 1's VXBs in sequence to perform the MVM of the convolution operator, passing its output to OP 2.
The traditional approach waits four cycles for OP 2 to begin its computation with $S_20$.(\fig\ref{fig:p3_xb} (c).)
In our MVM-grained pipeline, after OP 1 takes two cycles, OP 2 can start running on VXB 2,3 and then run on VXB4,5 after the next cycle, as shown in \fig\ref{fig:p3_xb} (d).
At most, four VXBs are activated simultaneously in contrast to that 6 VXBs in \fig\ref{fig:p3_xb} (c), reducing the peak power by $\sim$30\%.
Meanwhile, OP 2's inputs, $S_20_-0$ and $S_20_-1$, are half the size of $S_20$ in the traditional pipeline. Thus, the communication overhead in each computing stage is reduced, alleviating the pressure on on-chip bandwidth and the risk of pipeline stall.

We obtain the updated duplication number and more compact pipeline in this optimization grained and we will pass the duplication result to the next optimization grained.

\begin{figure}[tb]
    \centering
    \includegraphics[width=\linewidth]{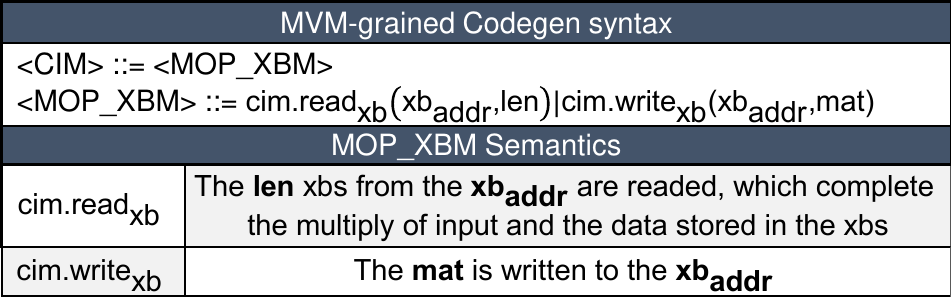}
    \caption{The syntax of MVM-grained codegen in BNF format and MOP\_XBM semantics.}
    \label{fig:p3_syntax_xb}
\end{figure}

\textbf{Meta-operator Flow Generation} 
Upon completing MVM-grained optimization, the compiler utilizes a meta-operator designed for XBM (MOP\_XBM) to describe the hardware activation at the core tier. 
The syntax at this grained is shown in the \fig \ref{fig:p3_syntax_all} and \fig \ref{fig:p3_syntax_xb}. Specifically, MOP\_XBM includes the $CIM.read_{crossbar}$ instruction for reading a specific crossbar to perform an MVM and $CIM.write_{crossbar}$ instruction for writing values like convolution weights to the crossbar.

\subsubsection{VVM-Grained Optimization}

\begin{figure}
    \centering
    \includegraphics[width=\linewidth]{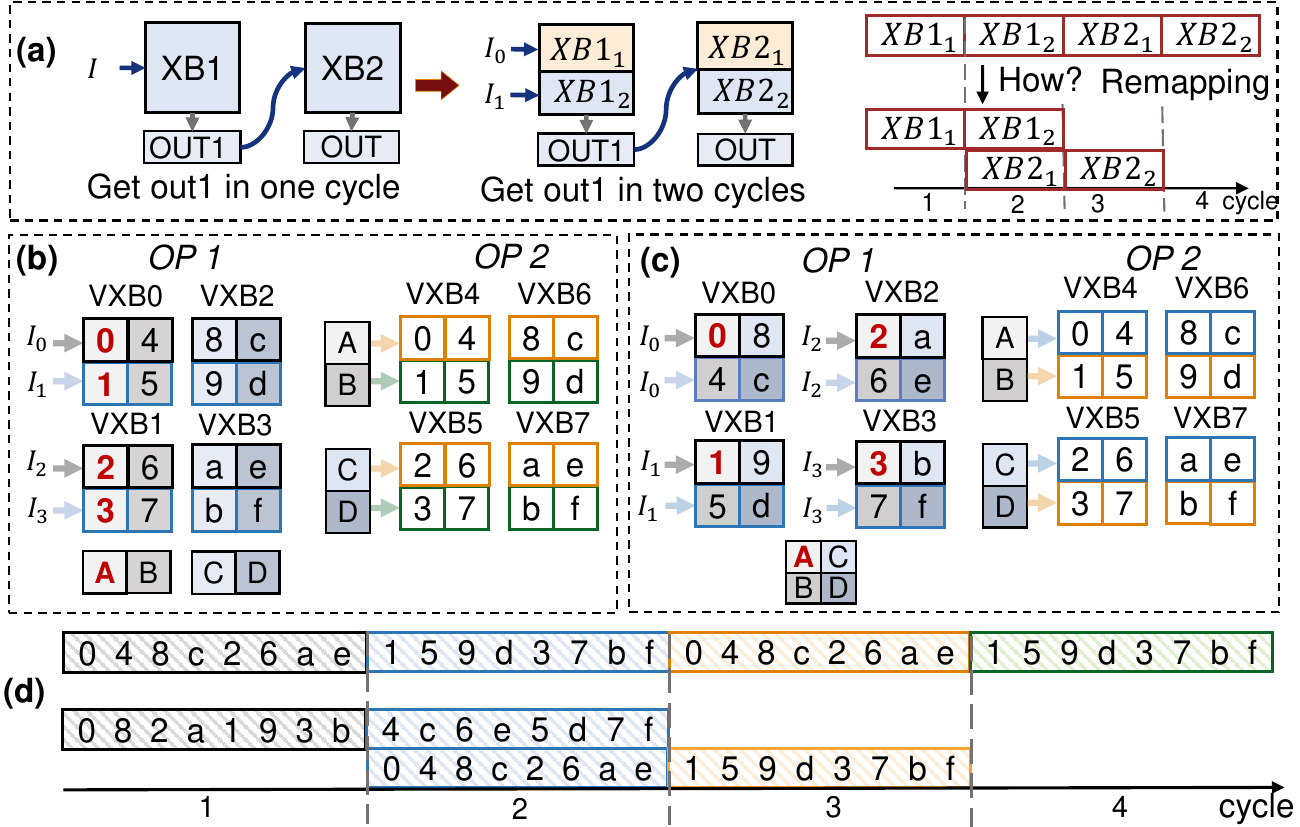}
    \caption{VVM-Grained Optimization (a) Parallelism opportunities in WLM (b) Na\"{\i}ve data mapping (c) Data remapping strategy (d) VVM-grained pipeline with \naive data mapping (up) and with our proposed data remapping (down).}
    \label{fig:p3_wl}
\end{figure}

In the WLM, partial rows within the crossbar can be activated at once, providing a more compact interface for vector-matrix multiplication compared to XBM which activates a whole crossbar for one computation.
As shown in the \fig \ref{fig:p3_wl}(a), when coarse-grained operators can be decomposed into fine-grained ones, we can explore additional parallelism opportunities.
To further improve computing throughput, we propose an innovative data \textit{remapping} strategy to enable a finer pipeline for the WLM CIM, which accounts for the updated computing graph with operator duplication results and whole three-tier abstraction.

The main idea of the remapping strategy is to distribute the data that contributes to the same computation to different crossbars. The proposed remapping strategy is illustrated in \fig\ref{fig:p3_wl}, where OP 1 and OP 2 are two adjacent operators in DNN and the output of OP 1 will be the input of OP 2.
For the example in \fig\ref{fig:p3_wl} (b), \textbf{\texttt{parallel row}} is half of the \textbf{\texttt{xb\_size}} rows, which means only row 0 and row 4 in VXB 0 can be activated in one cycle.
In the \naive data mapping strategy (\fig\ref{fig:p3_wl} (b)), since OP 1's output A is the accumulation of output of rows 0, 1, 2, and 3, it needs two cycles to get the output A (\fig\ref{fig:p3_wl} (d) up).
OP 2 can not start its computing until outputting A is complete.
As a result, OP 2 has to delay its computation for one cycle.
In contrast, our data remapping scheme maps rows 1 and 3 to different VXBs, as shown in \fig\ref{fig:p3_wl} (c). Then, rows 0-3 can complete their computations and accumulate to A in one cycle(\fig\ref{fig:p3_wl} (d) down).
The OP 2 can start its computation at Cycle 2. Meanwhile, rows 4-7 can perform the computations and accumulate their results to get output B.
Therefore, as shown in \fig\ref{fig:p3_wl} (d), a pipeline with a higher throughput can be achieved using our remapping scheme.

\begin{figure}[tb]
    \centering
    \includegraphics[width=\linewidth]{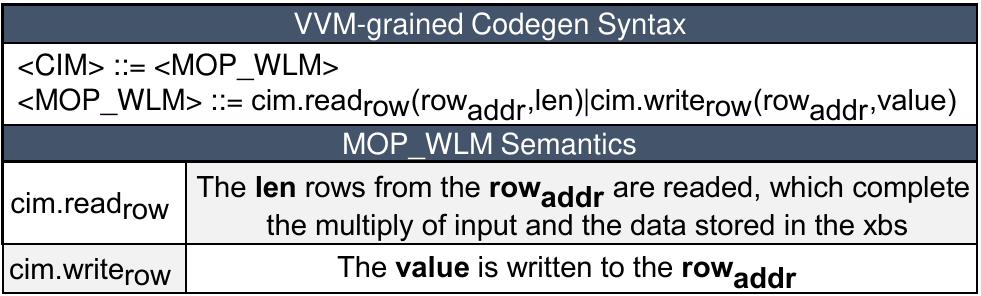}
    \caption{The syntax of VVM-grained codegen in BNF format and MOP\_WLM semantics.}
    \label{fig:p3_syntax_wl}
\end{figure}

\textbf{Meta-operator Flow Generation} 
Upon completing VVM-grained optimization, the compiler uses the meta operators specific for WLM (MOP\_WLM) to describe the corresponding hardware activation at the crossbar tier. As the \fig \ref{fig:p3_syntax_wl} shows, MOP\_WLM includes the $CIM.read_{row}$ instruction to read rows and $CIM.write_{row}$ instruction to write certain values to the rows.
CIM-MLC generates the meta-operator flow by invoking the MOP\_WLM, DCOM, and DMOV, for which the syntax is shown in the \fig \ref{fig:p3_syntax_all} and \fig \ref{fig:p3_syntax_wl}.

\begin{figure*}[tb]
    \centering
    \includegraphics[width=\linewidth]{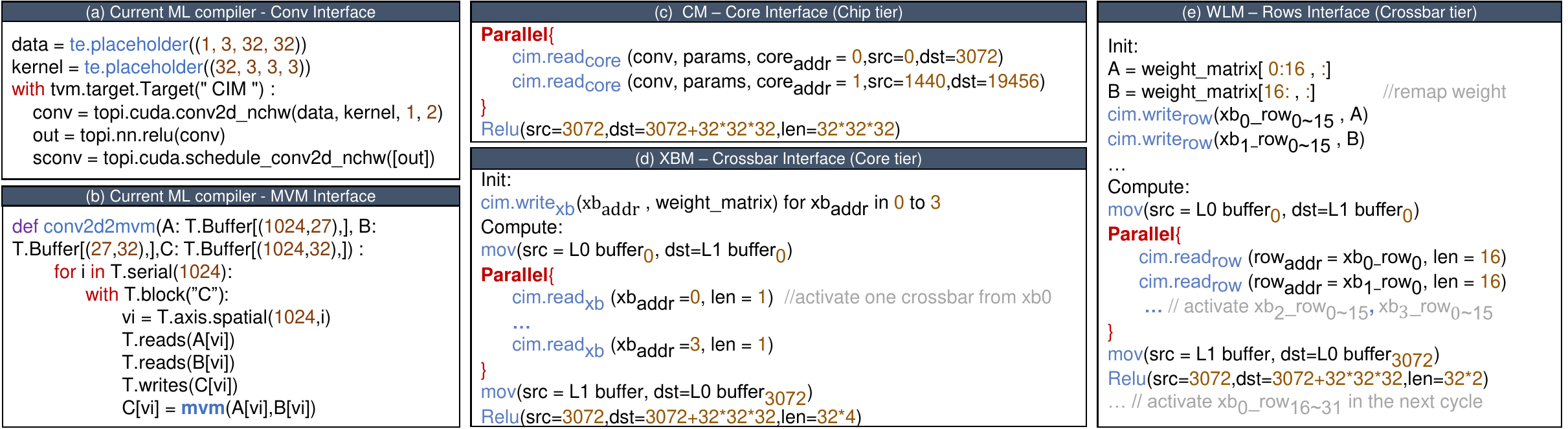}
    \caption{Generated code example for the Convolution-Relu. Left: Traditional DNN compilers; Right: CIM-MLC.}
    \vspace{-12pt}
    \label{fig:codegen}
\end{figure*}
\subsection{Putting it all together}
\begin{table}[!tbp]
\caption{Architecture Parameters of CIM example.}
\label{tab:example}
\resizebox{\linewidth}{!}
{
\begin{tabular}{@{}ll|ll|ll@{}}
\toprule
\multicolumn{2}{c|}{\textbf{Chip\_tier}} &\multicolumn{2}{|c|}{\textbf{Core\_tier}} & \multicolumn{2}{|c}{\textbf{Crossbar\_tier}}\\ \midrule
\begin{tabular}[c]{@{}c@{}}\textbf{core\_number} \end{tabular} & [2*1] & \begin{tabular}[c]{@{}c@{}}\textbf{xb\_number}\end{tabular} & [2*1] & \begin{tabular}[c]{@{}c@{}}\textbf{xb\_size}\end{tabular} & [32*128] \\
 & &  &   & \begin{tabular}[c]{@{}c@{}}\textbf{parallel row}\end{tabular} & 16\\
 \begin{tabular}[c]{@{}c@{}}\end{tabular} & &&  & \textbf{Precision} & 2-bit\\
\bottomrule
\end{tabular}}
\end{table}

In this section, we put the CIM-MLC compilation process all together. 
To help enhance users' understanding of our hardware abstraction and optimization methods, we have employed simplified networks and the CIM architecture to illustrate our overall compilation process.
We extracted commonly used operation in DNN, Convolution-Relu~\cite{agarap2018deep}, as an example. The parameters of the convolution are: input size:(3,32,32), kernel size:(32,3,3,3), stride:1, padding:1 with 8-bit precision for weight.
We assume that the target CIM architecture has 2 cores, each housing 2 crossbars with 32 row $\times$ 128 column memory cells. Each cell is capable of mapping 2 bits. As we take a portion of the compilation process from the complete network as an example, we simplify this architecture to support all common digital arithmetic operations, and the buffer bandwidths are ample, imposing no memory access limitations on the computation process. We use shared memory communication as NoC example.
The main architecture abstraction parameters are shown in the Table~\ref{tab:example}.
We will illustrate the generated code when assuming the architecture provides activation interfaces for cores, crossbars, and rows, separately.

\textbf{CG-Grained} When this architecture only provides activation interfaces for cores, which means CM, we apply CG-grained optimizations.

In this stage, we duplicate operators based on the hardware resource constraint. In this example, \textbf{\texttt{core\_number}} is 2 and each core can support the convolution with a kernel size of (32x3x3x3). 
Consequently, CIM-MLC decides the operator can be duplicated twice.

The final compiled meta-operator flow is illustrated in \fig \ref{fig:codegen}(c) CM, which sequentially completes the convolution and ReLU operations. 
To enable the parallel calculation of the duplicated operators, we partitioned the input feature maps into several sub-ones with the same number as the duplicated operators.
Then, we could get the buffer address of each sub-feature map. 
The duplication number for the operator is 2 and the feature map is partitioned into two sub-ones. So, we use two $cim.read_{core}$ to complete the convolution operation. As the two $cim.read_{core}$ have the same weights, their parameters are the same except for the $src$ and $des$ values. We use parallel\{$cim.read_{core}$(conv,params,0,0,3072), $cim.read_{core}$(conv,params,1,1440,19456)\} to denote executing computation on core 0 and core 1 in parallel. 
Furthermore, as we assume the architecture supports the ReLU operation, we can directly invoke the Relu meta-operator. 
So far, we have gotten the meta-operator flow for performing conv-relu operation on CM CIM by MOP-CM, DCOM and DMOV.

Traditional DNN compilers view this architecture merely as hardware for handling convolution and ReLU computations \cite{li2020deep}. Therefore, during the CG level optimization, they may directly invoke these interfaces to perform computation or merge the operators to reduce memory access. The generated code may be like \fig \ref{fig:codegen} (a) \cite{chen2018tvm}.
Compared to our method, they do not account for CIM-specific characteristics such as duplication opportunities during the mapping phase, which leads to a restricted exploration of the optimization possibilities.

\textbf{MVM-Grained} When this architecture provides programming interfaces at the core tier, which means XBM, we can perform the corresponding optimization, MVM-Grained optimization. In this stage, we explore the operator duplication within a core after converting the convolution to MVM. Given that each core has two crossbars, our approach allows us to update the operator duplication from 2 to 4 as each crossbar can support an MVM. 

As shown in the \fig \ref{fig:codegen}(d) XBM, after determining the updated duplication number, we first write the weight matrix to corresponding crossbars by $cim.write_{xb}$ before computing. Then, we load the data into local buffers using $mov$ operations and activate the four crossbars through the $cim.read_{xb}$ to complete four MVM operations in parallel. Since 1024 MVM operations are needed for one convolution, we generate 256 similar code segments to execute the entire convolution computation. 
Upon completing a batch of MVM operations, we perform ReLU calculations on the output of MVM to facilitate subsequent pipelines.

As we mentioned earlier in Section 1, traditional DNN compilation may be difficult to explore compilation optimizations suitable for CIM at MVM-grained. They usually focus on operators split and rearranged by loop unrolling \etc from the perspective of tensor size, but they cannot take advantage of the computing opportunity provided by the memory \cite{li2020deep}. The generated code is similar to the one depicted in \fig \ref{fig:codegen} (b) \cite{chen2018tvm}. For example, if we want TVM \cite{chen2018tvm} to support the optimization at MVM-grained for CIM, we must first register plenty of templates for various CIM architectures. Then, we need to modify the whole optimization pass of TVM, register the function interfaces supported by CIM, add new classes to expose the resources of the storage unit, and train a new automatic compilation optimization module. It is time-consuming and may completely destroy the existing compilation framework of TVM, which is not worth the loss.

\textbf{VVM-Grained} When this architecture offers rows activation interface within the crossbar, which means WLM, we will extend the data remapping at VVM-grained optimization, building upon the updated duplication.

Within the VVM optimization, we perform fine-grained control over the remapping of the weight matrix onto the crossbars. As the \textbf{\texttt{xb\_size}} row is 32 and \textbf{\texttt{parallel row}} is 16,  the original data that mapped to rows 0 to 15 (designated as `A') and rows 16 to 31 (designated as `B') in crossbar 0, is now divided into two distinct crossbars. `A' is remapped to crossbar 0’s rows 0 to 15, while `B' is remapped to crossbar 1’s rows 0 to 15 by the two $cim.write_{row}$ instructions.
This data remapping enables the concurrent computation of `A' and `B' within a single cycle by $cim.read_{rows}$ in parallel, a marked improvement over the previous mapping where `A' and `B' required separate activation in two cycles to perform the cumulative calculation essential for a single MVM computation. 
This simultaneous calculation of `A' and `B' within a single cycle facilitates subsequent pipelines.
Similar to code generation in the XBM, the compiler generated the meta-operators flow as shown in the \fig\ref{fig:codegen}(e) WLM. A total of 512 similar compute blocks are needed to complete the convolution.

\section{Experiment}
This section presents the evaluation of \names. 
We first present the implementation method of \names, and then conduct a comprehensive comparison and evaluation of \name with different CIM designs.

\subsection{Experiment Setup}

\textbf{Simulator} 
We developed a Python-based CIM functional simulator to verify the scheduling results, \ie the meta-operator execution trace. Additionally, we expanded upon an open-source simulator from previous studies~\cite{ankit2019puma,chen2018neurosim,dong2012nvsim} to evaluate the execution latency and power efficiency of the generated scheduling results. 

We verify the effectiveness of the functional simulator by comparing it with the Pytorch framework \cite{paszke2019pytorch}.
In our built functional simulator, the hardware abstraction of CIM is described by a data structure,
and meta-operators are implemented by specific functions. 
In this way, our functional simulator can perform the meta-operator flows as the DNN execution trace in the CIM.

We extended the open-source simulators proposed in previous works \cite{ankit2019puma,chen2018neurosim,dong2012nvsim} as the performance simulator to support the execution cycle and power consumption evaluation of meta-operators flow on the CIM-based DNN accelerators.
The primary extensions include: 1. We developed computational functions to facilitate simulating meta-operation execution, allowing the simulator to represent CIM architectures with varying computation granularity levels. 2. We establish the latency model, including computation, data movement, \etc~to evaluate the overall DNN latency.
\begin{table}[!tbp]
\caption{Architecture Parameters of CIM Hardware Baseline.}
\label{table1}
\resizebox{\linewidth}{!}
{
\begin{tabular}{@{}ll|ll|ll@{}}
\toprule
\multicolumn{2}{c|}{\textbf{Chip\_tier}} &\multicolumn{2}{|c|}{\textbf{Core\_tier}} & \multicolumn{2}{|c}{\textbf{Crossbar\_tier}}\\ \midrule
\begin{tabular}[c]{@{}c@{}}\textbf{core\_number} \end{tabular} & 768 & \begin{tabular}[c]{@{}c@{}}\textbf{xb\_number}\end{tabular} & 16 & \begin{tabular}[c]{@{}c@{}}\textbf{xb\_size}\end{tabular} & [128, 128] \\
\textbf{ALU (ops/cycle)} & 1024 & \textbf{ALU (ops/cycle)} & 1024  & \begin{tabular}[c]{@{}c@{}}\textbf{parallel row}\end{tabular} & 8\\
\begin{tabular}[c]{@{}c@{}}\textbf{L0\_BW}\end{tabular} & 384 b & \begin{tabular}[c]{@{}c@{}}\textbf{L1\_BW}\end{tabular} & 8192 b & \begin{tabular}[c]{@{}c@{}}\textbf{DAC/ADC}\end{tabular} & 1/8-bit \\
 \begin{tabular}[c]{@{}c@{}}\end{tabular} & &&  & \textbf{Type/Precision} & RRAM/2-bit\\
\bottomrule
\end{tabular}}
\end{table}

\textbf{CIM Architecture Baseline}
We refer to ISAAC \cite{Shafiee2016isaac} to establish a CIM architecture and use it as the baseline. The parameters of the baseline architecture are listed in Table~\ref{table1}. The parameters that are not elaborated are considered ideal, 
indicating that their influence on the evaluation is disregarded. For example, if the on-chip buffer is assumed to have sufficient bandwidth, load/store time can be hidden within the computation time.
We evaluate our \name on the baseline to verify its effectiveness and compare it with Poly-Schedule~\cite{han2021polyhedral}, which is also a compilation work for CIMs. 

\begin{figure}[!tb]
    \centering
    \includegraphics[width=1\linewidth]{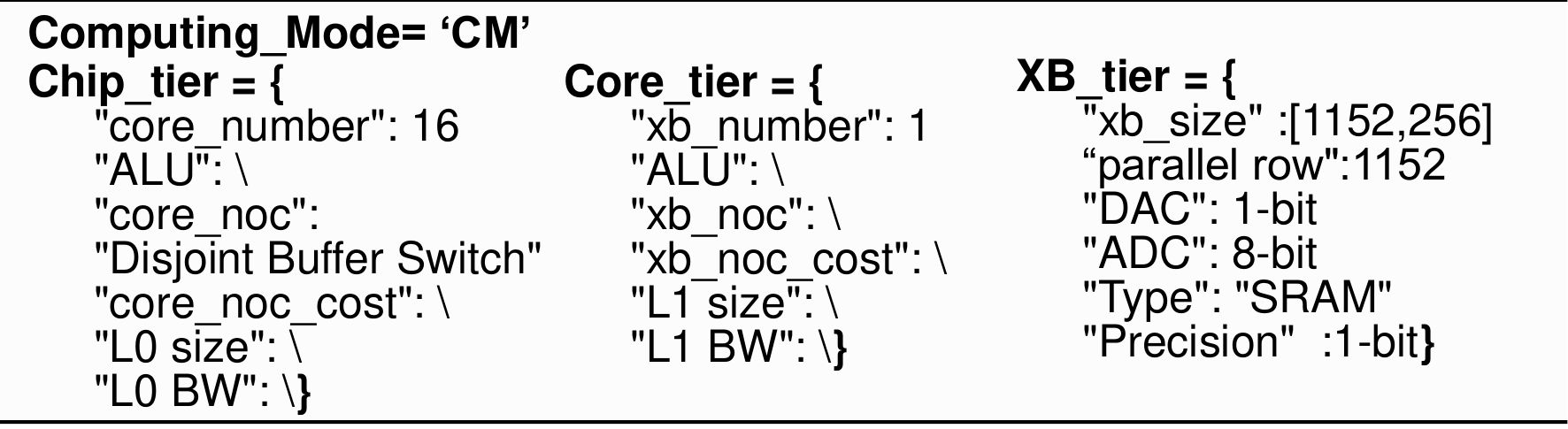}
    \caption{Architecture Abstraction of Jia \etal's work~\cite{jia202115}.}
    \label{cm}
\end{figure}
\begin{figure}[!tb]
    \centering
    \includegraphics[width=1\linewidth]{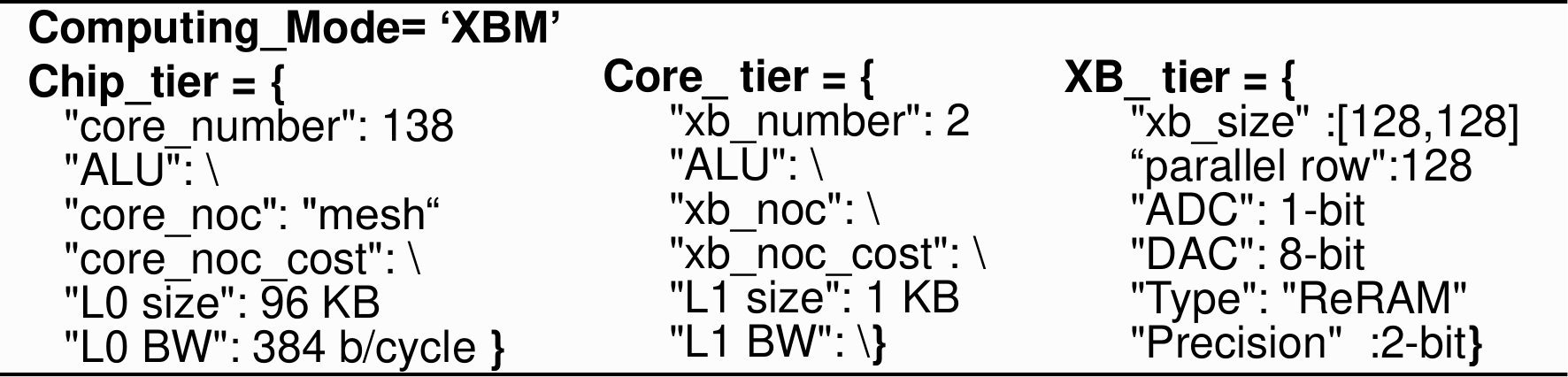}
    \caption{Architecture Abstraction of PUMA~\cite{ankit2019puma}.}
    \label{xbm}
\end{figure}
\begin{figure}[!tb]
    \centering
    \includegraphics[width=1\linewidth]{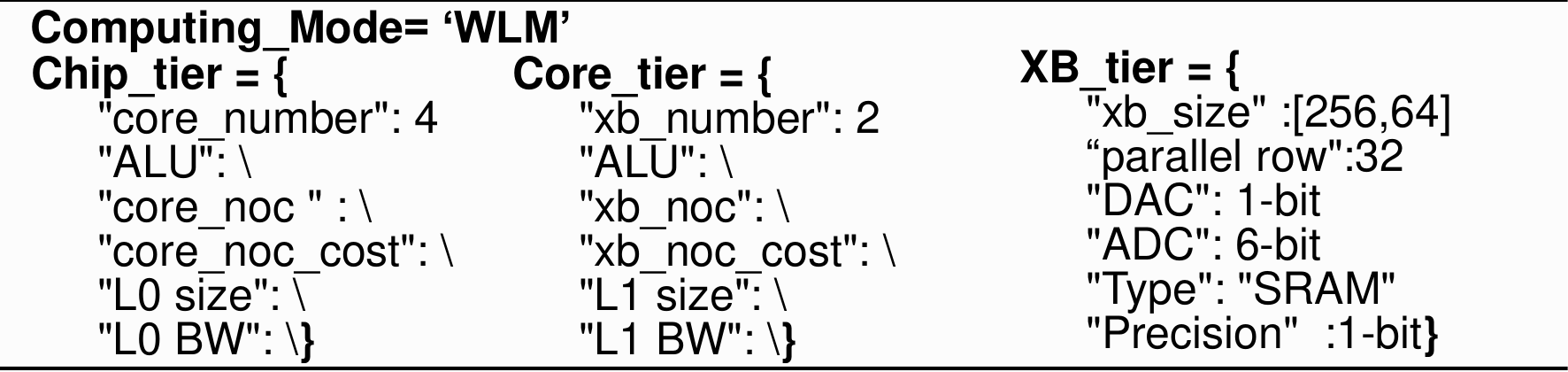}
    \caption{Architecture Abstraction of Jain \etal's~work~\cite{jain2021wlm}.}
    \label{wlm}
\end{figure}

\textbf{Network Benchmark}
First, to verify the generality of the scheduling method for different neural network tasks, we tested our operator scheduling method on multiple classic network models, including the VGG series~\cite{simonyan2014very}, ResNet series~\cite{he2016deep}, visual transformer (ViT)~\cite{dosovitskiy2020image}, \etc. All models’ weights and activation values are quantized with 8-bit precision and are tested on the ImageNet dataset.

\begin{figure*}[!tbp]
\centering
\includegraphics[width=\linewidth]{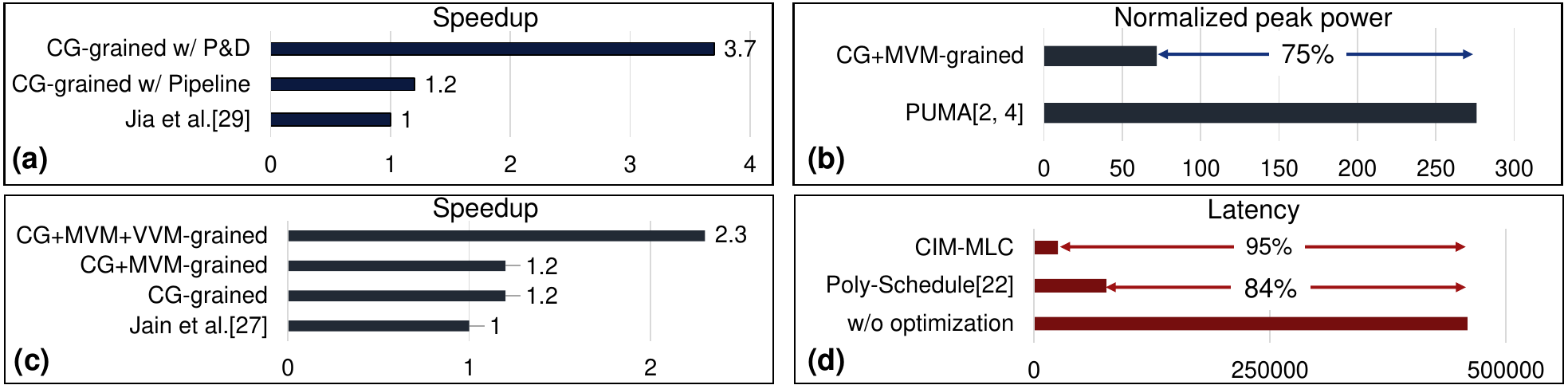}
\caption{(a) Comparison of the speedup between this work and the schedule method in work \cite{jia202115}; (b) Comparison of the peak power consumption between this work and the schedule method in PUMA \cite{ambrosi2018hardware,ankit2019puma}; (c) Comparison of the speedup between this work and the schedule method in work\cite{jain2021wlm}; (d) Comparison of the latency between this work and the schedule method in work \cite{han2021polyhedral}.}
\label{fig:ex5}
\end{figure*}

\textbf{Hardware Benchmark} To verify the generality of CIM-MLC for different CIM architectures, we conduct experiments on three CIM-based accelerators~\cite{ankit2019puma,jain2021wlm,jia202115}with different device types, precision, architecture hierarchy, and programming interfaces. Among them, Jia~\etal~\cite{jia202115} proposed an SRAM-based CIM accelerator that includes 16 CIMUs with a size of 1152x256, which incorporates embedded digital logic and high-precision ADC to achieve parallel activation calculation of 1152 rows. PUMA~\cite{ankit2019puma} is a programmable CIM architecture based on ReRAM to support a wide range of neural network applications. Jain \etal~\cite{jain2021wlm} introduces a CIM SRAM macro, in which only limited rows ($\leq$ 32) can be activated simultaneously in a crossbar to alleviate the computing variation.

\subsection{Effectiveness of CIM-MLC}

In this section, we apply the CIM-MLC in existing CIM-based accelerators~\cite{ankit2019puma,jain2021wlm,jia202115} to verify the generality of CIM-MLC and also compare CIM-MLC with previous work to show the effectiveness of our proposed optimization.

Firstly, we demonstrate the comparison results of \name with these three different CIM-based accelerators. These works have not only the specific CIM architecture but also the performance optimization methods. So, we first verify our abstraction techniques on their CIM distinct architectures and then compare the optimized performance by our method with their performance.

\textbf{Work 1:}
The hardware abstraction results of Jia~\etal's work are shown in \fig\ref{cm}. The parameters that are not elaborated are considered ideal and denoted with "$\setminus$". The computing mode abstraction of this design is the core mode (CM).
\name will apply CG-grained optimization to generate the meta-operator flow when mapping and scheduling the DNN.
The performance comparison between \name and Jia \etal's work is shown in \fig\ref{fig:ex5} (a). The CG-grained P\&D is the combination of the dynamic programming-based duplication and the pipeline, bringing about $3.7\times$ speedup over Jia \etal's work as it can make full use of the limited resources and speed up the computing of the bottleneck layer in DNNs.
The pipeline strategy can only achieve the speedup by $1.2\times$ over Jia \etal's work because the model size exceeds on-chip resources and the performance would not be fully optimized without the data mapping design.

\textbf{Work 2:}
Our hardware abstraction results for PUMA~\cite{ankit2019puma} are shown in \fig\ref{xbm}, where the computing mode is XBM. 
Thus, \name can perform CG-grained and MVM-grained optimization. 
We compare PUMA with our work on the VGG16, and the results are shown in \fig\ref{fig:ex5} (b). 
When mapping the VGG16 model, the proposed MVM-grained optimization fully utilized the scheduling space of the XBM mode in PUMA architecture and introduced a fine-grained MVM-grained pipeline. 
Our evaluation of peak power includes the power consumption of ADC/DAC, XB activation computation, and data movement. Our assessment shows that these three parts account for 10\%, 83\%, and 7\%, respectively. Thus, CIM-MLC performs fine-grained time-division activation of XBs and associated ADC/DACs, thereby significantly reducing the peak power by 75\%.

\textbf{Work 3:} 
\fig\ref{wlm} presents the hardware abstraction of Jain \etal's CIM macro~\cite{jain2021wlm} that has $WLM$ computing mode.
For a fair comparison, we use the VGG7 model as the benchmark and evaluate the scheduling results of \name and the original result under the same resource constraints. As shown in \fig\ref{fig:ex5} (c), with the three-level scheduling optimization (\ie CG-grained, MVM-grained, and VVM-grained), we achieve a speedup of about $2.3\times$ over Jain \etal's work.
Meanwhile, we show the speedup of CG-grained optimization and MVM-grained optimization, respectively. CG-grained optimization can achieve a speedup of $1.2\times$, while MVM-grained optimization cannot further bring effective acceleration improvement. This is because this CIM macro has limited on-chip resources, especially the small number of XBs in a Core, leading to the ineffectiveness of MVM-grained optimization for improving the speedup. The VVM-grained optimization fully improves the computing pipeline efficiency between adjacent operators by converting serial computations into parallel computations, thereby achieving a speedup.

\begin{figure*}[!tbp]
\centering
\includegraphics[width = \linewidth]{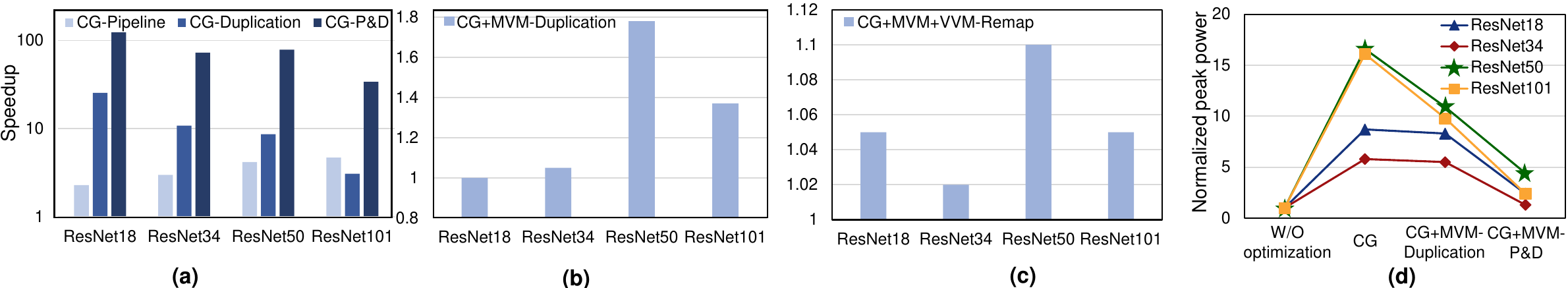}
\caption{ (a) Speedup of CG-grained optimization; (b) Speedup of CG+MVM-grained optimization; (c) Speedup of CG+MVM+VVM-grained optimization; (d) Comparison result of the peak power.}
\label{fig:ex2}
\end{figure*}

\textbf{Comparison to CIM-oriented compilers:}
We compared our method with the existing general-purpose CIM compiler tool Poly-Schedule~\cite{han2021polyhedral}. 
Poly-Schedule~\cite{han2021polyhedral} supports the compilation of CIM-based accelerators for CM and XBM, using operator duplication techniques based on greedy strategies and batch pipeline strategies for acceleration. 
Compared with this work, \name can optimize the internal computation pipeline of a single input image and explore the fine-grained scheduling space of the XBM mode to achieve better scheduling. 
We compared the operator scheduling results between Poly-Schedule and \name in the same CIM architecture abstracted in Table \ref{table1}. As shown in \fig\ref{fig:ex5} (d), compared to the latency result without any optimization, the Poly-Schedule~\cite{han2021polyhedral} utilizes on-chip resources with a greedy strategy to reduce 84\% computation cycles. Our work explores the fine-grained optimization space and reduces the computation cycles by up to 95\%, which achieves about $3.2\times$ speedup compared to the work \cite{han2021polyhedral}.

\subsection{Performance Analysis}
In this section, we analyze the speedup results of different granularity scheduling optimization in the multi-level scheduling for the baseline architecture in Table \ref{table1}. 
The network benchmark is the ResNet series~\cite{he2016deep}.
The results are shown in \fig\ref{fig:ex2}. In \fig\ref{fig:ex2} (a), we separate the optimization methods (\ie CG-Pipeline and CG-Duplication, CG-P\&D) in the CG-grained optimization and investigate their results, and the results are normalized to the results of the baseline architecture in CM mode without any optimization. In \fig\ref{fig:ex2} (b)-(c), the fine-grained optimization is with the coarse-grained ones to show the advantage of the multi-grained optimization. 
The results of \fig\ref{fig:ex2} (b) are normalized to the CG-P\&D results in  \fig\ref{fig:ex2} (a), and \fig\ref{fig:ex2} (c)'s results use the results in \fig\ref{fig:ex2} (b) as the baseline.

\begin{figure*}[!tbp]
\centering
\includegraphics[width=\linewidth]{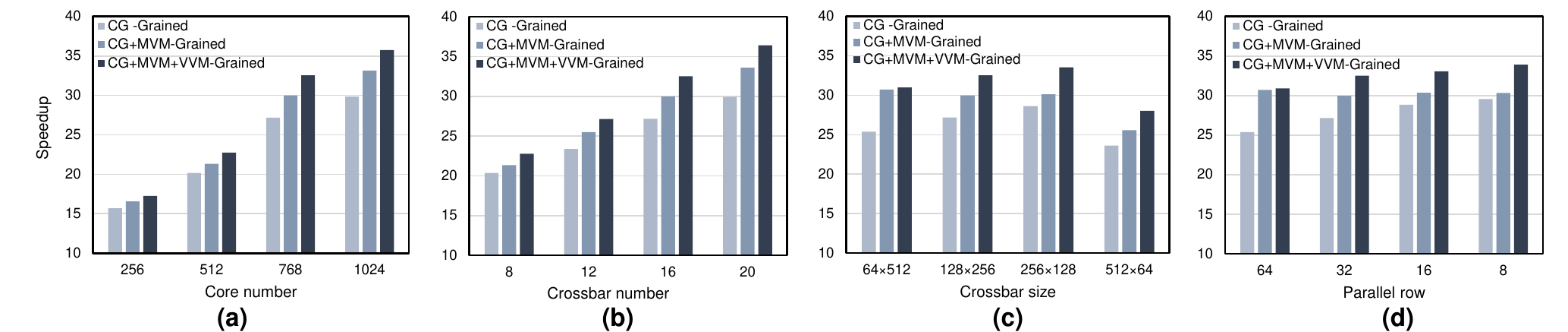}
\caption{ (a) Scheduling results with the different number of cores in a chip; (b) Scheduling results with different numbers of crossbars in a core; (c) Scheduling results with different crossbar sizes;
(d) Scheduling results with different parallel rows in a crossbar.}
\label{fig:ex4}
\end{figure*}

It can be observed that the model structure has a significant impact on the scheduling results. 
As shown in \fig\ref{fig:ex2} (a), in CG-grained optimization, as the depth of ResNet increases, the speedup achieved by pipeline (CG-Pipeline) increases from $2.3\times$ to $4.7\times$.
The pipeline strategy can fully perform parallel computation between adjacent operators. 
However, with the increase of the model size (from ResNet18 to ResNet101), the speedup of duplication (CG-Duplication) decreases from $25.4\times$ to $3.1\times$. 
When combining the pipeline and duplication strategies (CG-P\&D), ResNet series models achieve significant performance improvements up to $123\times$.

As shown in \fig\ref{fig:ex2} (b), CG+MVM-Duplication increase the speedup of ResNet50/ResNet101 by approximately $1.8\times$ / $1.4\times$ over the CG-P\&D.
Meanwhile, as shown in \fig\ref{fig:ex2} (d), the normalized peak power consumption increases by approximately $5\times$ to $16\times$ for the ResNet series in CG-grained optimization since the number of crossbars that work at the same time increases, leading to the high power consumption of CIM accelerator.
 Then, 
 MVM-grained optimization reduces peak power consumption by up to 85\% (ResNet101) thanks to its fine-grained pipeline strategy that lowers the peak activated crossbar number.

The results in \fig\ref{fig:ex2} (c) show that the speedup of ResNet50 in the VVM-grained optimization can be 10\% higher than that of the MVM-grained optimization because the data remapping strategy improves the pipeline throughput. 

\subsection{Sensitive Study of CIM Architecture}
In this section, we evaluate the effect of the change of CIM architecture parameters on the scheduling results of \names. We use a transformer network architecture, ViT~\cite{dosovitskiy2020image}, as the benchmark.
The baselined architecture uses the parameters from \tab~\ref{table1} except for the crossbar size is 128 $\times$ 256.

\subsubsection{Core Number \& Crossbar Number}
First, we investigate the effect of different core numbers on the effectiveness of \names.
The results are shown in \fig\ref{fig:ex4} (a). 
As the total number of cores increases from 256 to 1024, the speedup achieved by \name grows as well. Since on-chip resources gradually increase, \name can fully show the potential of the duplication strategy as well as the pipeline.  
Therefore, the speedup of the CG-grained optimization increased from $15\times$ to $30\times$. 
In the MVM-grained optimization, as the number of cores increased, the proposed method could duplicate a single operator to different crossbars to fully utilize resources, achieving a maximum speedup improvement of about $1.1\times$ compared to the CG-grained optimization. 
In the VVM-grained optimization, a finer pipeline reduced the equivalent activation times of the crossbar, further resulting in a speedup improvement of approximately $1.2\times$ compared to the CG-grained result.
\fig\ref{fig:ex4} (b) shows the performance speedup achieved by \name when the number of crossbars in each core varies. 
Similar to the results of the core number, the speedup grows as the crossbar number increases. 
\subsubsection{Crossbar Size}

\fig\ref{fig:ex4} (c) compares the speedup achieved by \name as the crossbar size changes. As the crossbar row size increases from 64 to 256, the CG-grained speedup gradually increases. This is because larger row sizes increase the amount of input data required by the crossbar, which puts pressure on the bandwidth and leads to longer computing latency. The CG-grained pipeline optimization can reduce data feeding pressure, resulting in better acceleration. 
Since the weight matrix size of ViT matches the crossbar size when it changes from ($64\times 512$) to ($256\times 128$), the MVM-grained can make full use of redundant crossbar resources to achieve the same acceleration.
The speedup of VVM-grained gradually increases because as the crossbar column size decreases, the same weight needs to be mapped on more crossbars in the horizontal direction. The VVM-grained remapping strategy can make full use of the advantages of these horizontal crossbars, which reduce the calculation delay in the crossbar, and achieve better acceleration.
Upon increasing the number of crossbar rows to 512, the speedup has considerably decreased. This can be attributed to the fact that ViT comprises numerous matrices with a row size of 768, necessitating two vertical crossbars ($512\times 64$) for mapping. Consequently, ViT needs more resources, requiring it to be segmented and mapped on the given CIM, ultimately leading to a decrease in speedup.

\subsubsection{Parallel Row}
When the number of parallel rows in the crossbar is changed, the scheduling result can be observed in \fig\ref{fig:ex4} (d). If the number of parallel rows is decreased, it may be challenging to achieve effective acceleration through MVM-grained scheduling, while the VVM-grained remapping strategy can mitigate the impact of this reduction on latency. Specifically, when the number of parallel rows is 8, VVM-grained scheduling can achieve an acceleration of approximately 20\% beyond the optimization results of MVM-grained scheduling.

\section{Conclusion}
We propose \names, a general compilation tool for the CIM architecture that consists of hardware and computation abstraction and a multi-level operator scheduling method. The hardware abstraction method, Abs-arch, provides a unified description of CIM architectures from chip to crossbar tier. Three computing mode abstractions are established to represent different granularity of the programming interface. A set of \mps is established to describe the computing process under different modes. Besides, the multi-level scheduling optimization method is proposed to explore the acceleration potential of the CIM architecture, from CG-grained to MVM-grained and VVM-grained, and output the corresponding meta-operator flow. 
Comprehensive experimental results show that \name has wide software and hardware adaptability.  
What's more, the proposed \name can serve as the middleware between neural network models and CIM hardware, reducing the design burden on experts in the fields of CIM architecture and neural network algorithms.

\section{Acknowledgments}
This work was supported by the National Natural Science Foundation of China under NSFC.62222411.

\bibliographystyle{plain}
\bibliography{main}

\end{document}